\documentclass[11pt]{article}
\pdfoutput=1
\usepackage[utf8]{inputenc}
\usepackage{dsfont}
\usepackage{IEEEtrantools}\usepackage{amsfonts}
\usepackage{amsmath}
\allowdisplaybreaks[4]        
\usepackage{amssymb}
\usepackage{upgreek}
\usepackage{euscript}     
\usepackage{braket}
\usepackage{starfont}
\usepackage{color,soul}         
\usepackage{tensor}        
\usepackage{amsthm}
\usepackage{graphicx}
\usepackage{IEEEtrantools}
\usepackage{slashed}
\usepackage{leftidx}
\usepackage{subfigure}
\usepackage{bbm}
\usepackage{bbold}
\definecolor{outerspace}{rgb}{0.25, 0.29, 0.3}
\definecolor{scarlet}{rgb}{1.0, 0.13, 0.0}
\usepackage[header,title,page,titletoc]{appendix}  
\definecolor{princetonorange}{rgb}{1.0, 0.56, 0.0}
\definecolor{WildStrawberry}{rgb}{1.0, 0.26, 0.64}
\definecolor{rossocorsa}{rgb}{0.83, 0.0, 0.0}
\definecolor{navyblue}{rgb}{0.0, 0.0, 0.5}
\usepackage[numbers,sort&compress]{natbib}  
\usepackage{float}
\usepackage[paper=letterpaper,margin=1in]{geometry}
\usepackage{mathtools}

\newcommand{\req}[1]{(\ref{#1})} %{Eq.\thinspace(\ref{#1})}  

\newcommand{\be}{\begin{equation}}
\newcommand{\ee}{\end{equation} }
\newcommand{\beqa}{\begin{eqnarray}}
\newcommand{\eeqa}{\end{eqnarray}}
\newcommand{\beqar}{\begin{eqnarray*}}
\newcommand{\eeqar}{\end{eqnarray*}}

\newcommand{\ssc}{\scriptscriptstyle}
\newcommand{\eg}{{\it e.g.,}\ }

\newcommand{\cO}{\mathcal{O}}

\usepackage[normalem]{ulem}

\makeatletter
\newcommand{\dal}{\mathop{\mathpalette\dal@\relax}}
\newcommand{\dal@}[2]{%
  \begingroup
  \sbox\z@{$\m@th#1\square$}%
  \dimen0=\fontdimen8
    \ifx#1\displaystyle\textfont\else
    \ifx#1\textstyle\textfont\else
    \ifx#1\scriptstyle\scriptfont\else
    \scriptscriptfont\fi\fi\fi3
  \makebox[\wd\z@]{%
    \hbox to \ht\z@{%
      \vrule width \dimen0
      \kern-\dimen0
      \vbox to \ht\z@{
        \hrule height \dimen0 width \ht\z@
        \vss
        \hrule height 2\dimen0
      }%
      \kern-2.5\dimen0
      \vrule width 2.5\dimen0
    }%
  }%
  \endgroup
}
\makeatother

\definecolor{MidnightBlue}{RGB}{5,5,150}

\usepackage{hyperref}
\hypersetup{
    colorlinks,
    citecolor=MidnightBlue,
    filecolor=black,
    linkcolor=black,
    urlcolor=MidnightBlue
}

\numberwithin{equation}{section}

\begin{document} 

\begin{titlepage}
\hfill \\
\begin{flushright}
\hfill{\tt CERN-TH-2022-076}
\end{flushright}
\begin{center}

\phantom{ }
\vspace{1.5cm}

{\bf \LARGE{Higher-curvature Gravities from Braneworlds\\ \vspace{10pt} and the Holographic c-theorem}}

\vskip 1cm

{\bf Pablo Bueno${}^{a}$, Roberto Emparan${}^{b,c}$, Quim Llorens${}^{c}$}
\vskip 0.6cm

\small{${}^{a}$\textit{CERN, Theoretical Physics Department,}}\\
\small{\textit{CH-1211 Geneva 23, Switzerland}}
\medskip

\small{${}^{b}$\textit{Instituci\'o Catalana de Recerca i Estudis Avançats (ICREA),}}\\
\small{\textit{Passeig Llu\'is Companys 23, E-08010
Barcelona, Spain}}
\medskip

${}^{c}$\textit{Departament de F\'isica Qu\`antica i Astrof\'isica, Institut de Ci\`encies del Cosmos,}\\ \textit{Universitat de
Barcelona, Mart\'i i Franqu\`es 1, E-08028 Barcelona, Spain}

\bigskip

\texttt{pablo.bueno-gomez@cern.ch}, 
\texttt{emparan@ub.edu}, \texttt{qllorens@icc.ub.edu}

\vskip 1cm
\begin{abstract}
\noindent 
We study the structure of the higher-curvature gravitational densities that are induced from holographic renormalization in AdS$_{d+1}$. In a braneworld construction, such densities define a $d$-dimensional higher-curvature gravitational theory on the brane, which in turn is dual to a $(d-1)$-dimensional CFT living at its boundary.
We show that this CFT$_{d-1}$ satisfies a holographic $c$-theorem in general dimensions (different than the $g$-theorem of holographic boundary CFTs),
since at each and every order the higher-curvature densities satisfy $c$-theorems on their own. We find that, in these densities, the terms that affect the monotonicity of the holographic $c$-function are algebraic in the curvature, and do not involve covariant derivatives of the Riemann tensor. We examine various other features of the holographically induced higher-curvature densities, such as the presence of reduced-order traced equations, and their connection to Born-Infeld-type gravitational Lagrangians.  
\end{abstract}
\end{center}
\end{titlepage}

\setcounter{tocdepth}{2}

{\parskip = .2\baselineskip \tableofcontents}

\newpage

%%%%%%%%%%%%%%%%%%%%%%%%%%%%%%%%%%%%%%%%%%%%%%%%%%%%%%%%%%%%%%%%%

\section{Introduction and Summary}\label{sec:Introduction}

The quantum fluctuations of a field in a curved spacetime give rise to ultraviolet divergences that take the form of invariants of the metric and curvature in the quantum effective action. For holographic conformal field theories dual to Anti-de~Sitter spacetime in $d+1$ dimensions with radius $\ell$, the form of this action is \cite{deHaro:2000vlm,Balasubramanian:1999re,Emparan:1999pm}
\begin{align}\label{divact}
    I_{\text{div}}=\frac{\ell}{16\pi G_N(d-2)}\int_{\partial\mathcal{M}} d^d x\sqrt{-g}\biggl[
    &\frac{2(d-1)(d-2)}{\ell^2}
    +R\nonumber\\
    &+\frac{\ell^2}{(d-2)(d-4)}\left(R_{ab}R^{ab}-\frac{d}{4(d-1)}R^2\right)
    +\dots
    \biggr].
\end{align}
Here $g_{ab}$ is the metric induced near the AdS boundary $\partial\mathcal{M}$, and the divergences arise because $g_{ab}$ grows infinitely large as the asymptotic boundary is approached. After regularization, counterterms are added with the same structure as \eqref{divact} in order to renormalize the theory.

The effective action expansion in \eqref{divact} can be systematically derived from the bulk Einstein equations in asymptotically AdS spacetimes \cite{Kraus:1999di,Papadimitriou:2004ap,Papadimitriou:2010as, Papadimitriou:2016yit,Elvang:2016tzz}, and we will give the explicit results up to quintic order for general dimension $d$, and to sextic order for $d=3$. The coefficients of each of the individual curvature invariants reflect the ultraviolet structure of holographic CFTs,\footnote{Even though it is not known whether non-trivial CFTs exist in arbitrary $d$, holography suggests that their leading planar limit exists (at least for generalized free fields).} and although they have been known for many years, their specific form appears to have received little attention. In this article we will investigate some of their properties from a point of view that directly connects them to (i) higher-curvature theories of gravity, and (ii) holographic $c$-theorems.

\begin{figure}%[th]
        \centering
         \includegraphics[width=.3\textwidth]{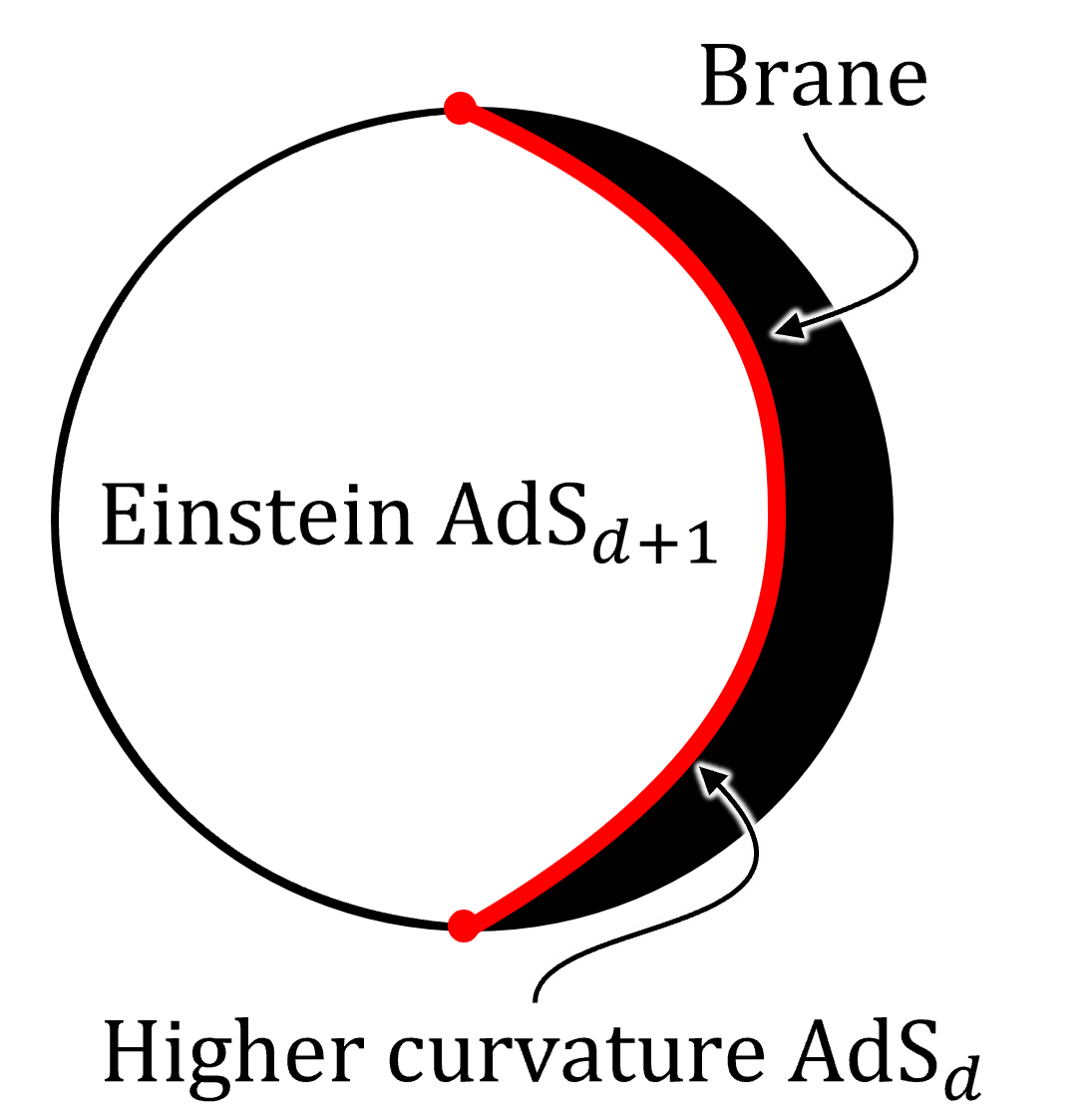}
        \caption{\small Braneworld gravity and holography. The bulk is described by Einstein-AdS$_{d+1}$ gravity. The black region is excluded by the introduction of a brane, where a gravitational theory with higher-curvature terms is induced. When the brane geometry is asymptotically AdS$_d$ (as in the figure), this higher-curvature gravitational theory can be dualized to a CFT$_{d-1}$ at its boundary (red dots). This leads to a doubly-holographic construction of boundary CFT, but this view will not be prominent in our article, where we regard the higher-curvature theory (and its dual CFT$_{d-1}$) on its own, regardless of its coupling to a CFT$_d$ dual to the AdS$_{d+1}$ bulk.}
        \label{fig:braneworld}
\end{figure}

\paragraph{Holographically induced higher-curvature gravity.} For this purpose, we will introduce a brane near the boundary of the AdS bulk, as in a Randall-Sundrum braneworld construction \cite{Randall:1999vf}. The brane effectively acts as a cutoff that renders the action  \eqref{divact} finite, and furthermore, it makes the metric $g_{ab}$ dynamical. Then, \eqref{divact} is interpreted as the effective action of the gravitational theory that is induced on the $d$-dimensional brane, with a Newton's constant $G_{\rm{eff}}=(d-2)G_N/\ell$, and with the brane tension adding to the cosmological constant term \cite{deHaro:2000wj}.\footnote{See \eg, \cite{Chen:2020uac} for more details. If we consider the brane to be two-sided, then \eqref{divact} will contribute twice to the effective action. Since we are only interested in the structure of the curvature terms, these considerations will be immaterial for us.}  In effect, the Einstein-Hilbert term and all the higher-curvature operators in the effective action are generated when the bulk Einstein equations are solved in the region near the boundary excluded by the introduction of the brane. In dual terms, gravitational dynamics is induced from the integration of the ultraviolet degrees of freedom of the CFT above the cutoff. As a result, we obtain a holographic realization of `induced gravity' (figure~\ref{fig:braneworld}).

In this manner, we can view the braneworld construction as a means of generating a specific theory of higher-curvature gravity. The $d$-dimensional action must be regarded as an effective theory with an infinite series of terms, each naturally smaller than the previous one. Since the $(d+1)$-dimensional Einstein bulk theory is well defined, we expect that this good behavior is inherited by the $d$-dimensional effective theory---at least for the entire series. However, one may also attempt to truncate the expansion at a finite order, and hope that the higher-curvature gravitational theory that results is, if not completely well-defined by itself, at least special in some respects. That is, we are proposing the holographic braneworld perspective as an appealing rationale motivating a class of higher-curvature theories with distinctive properties, which we shall investigate in this article.

\paragraph{Holographic $c$-theorem for induced higher-curvature gravities.} The braneworld construction can also have another ramification that we will exploit. In general, depending on the value of the brane tension, the cosmological constant that is induced on the brane theory can be positive, negative or zero. The three cases give valid higher-curvature effective theories, but when the cosmological constant is negative, and the geometry on the brane is asymptotically AdS$_d$ (known as a Karch-Randall braneworld \cite{Karch:2000ct}), we can perform one more holographic dualization. Namely, we can envisage that the gravitational theory on the brane is itself dual to a CFT$_{d-1}$ at its boundary. 

The usual interpretation of this doubly-holographic setup is in terms of duality to a boundary CFT, that is, a CFT$_{d}$ in a space with a boundary where a CFT$_{d-1}$ lives \cite{Karch:2000ct,Takayanagi:2011zk}. However, this view will not play a role in this article. Once we have obtained a gravitational theory on AdS$_d$, we will be considering it on its own, without regard to its possible coupling to the holographic CFT$_d$. Then, the CFT$_{d-1}$ to which our gravitational theory is dual will be different than the one that resides at the boundary of the CFT$_{d}$ in doubly-holographic setups. In other words, for us the holographic construction is simply a means of generating a specific class of higher-curvature gravitational theories which are plausibly dual to conformal field theories, but these are not necessarily coupled to any other system.

We will prove that these holographic CFT$_{d-1}$ possess a basic property of well-defined conformal theories, namely, they satisfy $c$-theorems. Holographic theories incorporate renormalization group flows as bulk solutions that interpolate between two asymptotically AdS regions \cite{Freedman:1999gp,Girardello:1998pd}. These act as the UV and IR fixed points, while the bulk radial coordinate parametrizes the flow.  Holographically, one expects that the $c$-function should be a rough measure of the curvature radius of the geometry, such that it monotonously decreases along the flow from the boundary into the bulk.

We will actually find a stronger result: the higher-curvature theories that are defined by the Lagrangian densities at each order in the expansion \eqref{divact} separately satisfy  holographic $c$-theorems. Although this might not be unexpected given the good behavior of the `parent theory' that gives rise to them, it is not a direct consequence of the $c$-theorem of the holographic CFT$_{d}$. Neither is it the same as the $g$-theorem for holographic boundary CFTs in \cite{Fujita:2011fp} since, as we mentioned above and will discuss later in more detail, our CFT$_{d-1}$ are differently defined, and our method of proof and bulk interpretation of the result are also very different.

\paragraph{Other properties of higher-curvature densities.}

The proof of these holographic $c$-theorems relies on particularities of the $d$-dimensional order-$n$ densities, but not in a very detailed way. 
Further examination of their structure, up to the highest order we have computed them, reveals finer features. In particular, we can decompose each order-$n$ density $\mathcal{L}_{(n)}$ appearing in the brane effective action into a linear combination of a term $\mathcal{S}_n$, that gives a non-trivial $c$-function, and a term $T_n$ that does not contribute to it, since  it identically vanishes on the renormalization group flow geometry. We find evidence that this decomposition can always be made in such a way that all the $\mathcal{S}_n$ are algebraic in the Riemann tensor, with no derivatives of it. That is,
\begin{equation}\label{Lng}
     \mathcal{L}_{(n)} \propto \ \mathcal{S}_n[R_{abcd}]+T_n[R_{abcd},\nabla_a]\, .
\end{equation}
We have proven that this is possible for all $n$ in $d=3$, and strong evidence suggests that it should hold for all $n$ and $d$.

Using the decomposition \eqref{Lng}, we have then looked for other special properties of these densities. In most cases, we do not have proofs that apply to all orders and dimensions, but instead we have identified particular features by direct inspection of the terms that we have explicitly generated.

A first observation follows directly from the form of the first three orders in the effective action, shown in \eqref{divact}. In any dimension $d$, we have
\begin{equation}
    T_0=T_1=T_2=0\,.
\end{equation}
In particular, in $d=3$ the only quadratic order term is, up to an overall factor,
\begin{equation}\label{3dS2}
    \mathcal{S}_2 = R_{ab}R^{ab} - \frac{3}{8} R^2\, ,
\end{equation}
which, as noted in \cite{Emparan:2020znc}, is the same density as in the New Massive Gravity (NMG) of \cite{Bergshoeff:2009hq}.
At the next, cubic order, the $T_n$ make appearance in every $d$ (see \req{cubicca} below). In $d=3$, up to an overall factor, we find
\begin{equation}\label{3dS3}
    \mathcal{S}_3 = R_{a}^b R_{b}^c R_c^a + \frac{17}{64} R^3 - \frac{9}{8} R R_{ab}R^{ab}\, ,
\end{equation}
and
\begin{equation}
    T_3=\frac{1}{2}C_{abc}C^{abc}\,,
\end{equation}
where $C_{abc}$ is the Cotton tensor. Both these densities have featured in earlier literature:  $\mathcal{S}_3$ was proposed in \cite{Sinha:2010ai} as a cubic generalization of NMG that satisfies a holographic $c$-theorem, and $T_3$ defines the only cubic theory whose equations of motion have a third-order trace \cite{Oliva:2010zd}.

The appearance of \eqref{3dS2} and \eqref{3dS3} might point to a stronger link between the three-dimensional massive gravity theories of  Karch-Randall braneworlds and the generalized higher-curvature theories that satisfy holographic $c$-theorems \cite{Paulos:2010ke}. Note, however, that the origin of the graviton mass in Karch-Randall braneworlds is tightly linked to its coupling to the dual CFT \cite{Neuenfeld:2021wbl}, which is in general absent in NMG and its generalizations.\footnote{Note also that the coefficient of the Einstein-Hilbert term in NMG is negative  \cite{Bergshoeff:2009hq}, opposite to the `normal' sign it has in the braneworld, as seen in \eqref{divact}.}

For general higher dimensions, the cubic densities $\mathcal{S}_3$ and $T_3$ are also special in similar ways. We find that $\mathcal{S}_3$ can be identified with a linear combination of the cubic Quasi-topological gravity density \cite{Oliva:2010zd,Oliva:2010eb,Myers:2010ru}, which has second-order traced equations, plus a density which contributes trivially to the $c$-theorem. On the other hand, $T_3$ turns out to be given by another previously identified combination \cite{Oliva:2010zd}, distinguished, just like in three dimensions, by possessing third-order traced equations.

The reduced-order property of the traced equations is a rather stringent feature, but in holographically induced gravities it does not seem to generally hold beyond cubic terms. Indeed, the quartic term $T_4$ already does not satisfy it in $d=3$. 

Finally, also in three dimensions, we have found an intriguing connection between the full tower of counterterms and the Born-Infeld-like extension of NMG presented in \cite{Gullu:2010pc}. At present, we do not know whether this finding is fortuitous, or instead it has a deeper meaning. 

\medskip

The remainder of the paper proceeds as follows. In section \ref{sec:Counterterms}, we review the computation of the effective action \eqref{divact} from holographic renormalization, and then expand it to quintic order in general dimensions and to sextic order in $d=3$. In section \ref{sec:HoloCTheorem} we review the holographic $c$-theorem construction for higher-curvature gravities, and also present a few new observations on the topic. Then, in section \ref{sec:GenProof}, we prove that all the terms in the effective action separately fulfill a holographic $c$-theorem. In sections \ref{sec:3DResults} and \ref{sec:GenD} we study the structure of each order-$n$ density in \eqref{divact}, in $d=3$ and in general dimensions, respectively. We end with comments on possible future directions.

%%%%%%%%%%%%%%%%%%%%%%%%%%%%%%%%%%%%%%%%%%%%%%%%%%%%%%%%%%%%%%%%%

\section{Holographic renormalization and induced gravity}\label{sec:Counterterms}

We begin with a sketch of how the action \eqref{divact} arises. The starting point is the gravitational bulk action for a $(d+1)$-dimensional asymptotically AdS spacetime,
\begin{equation}\label{bulkaction}
    I = \frac{1}{16 \pi G_N} \left[ \int_{\mathcal{M}} \ d^{d+1}x \sqrt{-G} \left( R[G] + \frac{d(d-1)}{\ell^2} \right) + 2 \int_{\partial\mathcal{M}} d^dx \sqrt{-g} K \right]\,.
\end{equation}
Near the asymptotic boundary, we write the bulk metric in a Fefferman-Graham expansion as \cite{Fefferman:2007rka}
\begin{align}
    %ds^2 = 
    G_{\mu\nu} dx^\mu dx^\nu &= \frac{\ell^2}{4\rho^2} d\rho^2 + \frac{\ell^2}{\rho} \hat{g}_{ij}(\rho,x) dx^i dx^j\nonumber\\
    &= \frac{\ell^2}{4\rho^2} d\rho^2 + \frac{\ell^2}{\rho}\left(\hat{g}^{(0)}_{ij}(x) + \cO (\rho)\right) dx^i dx^j\,.
\end{align}
We then solve the Einstein equations, order by order in $\rho$, in terms of the `renormalized metric' $\hat{g}^{(0)}_{ij}$ and its derivatives \cite{deHaro:2000vlm}. This series solution is then plugged back into the bulk action, and, after introducing a cutoff at $\rho=\varepsilon$, the bulk coordinate $\rho$ is integrated between $\varepsilon$ and a finite value of $\rho>\varepsilon$. The result is a series expansion where the first terms diverge as $\varepsilon\to 0$ in the form
\begin{align}\label{evend}
    I
    = \frac{\ell}{16\pi G_N} \int d^d x \sqrt{-\hat{g}^{(0)}} \left( \varepsilon^{-d/2} \hat{\mathcal{L}}_{(0)} +  \cdots + \varepsilon^{-1} \hat{\mathcal{L}}_{(\lceil d/2 \rceil -1)} - \log (\varepsilon) \hat{\mathcal{L}}_{(d/2)} \right) + \mathcal{O}(\varepsilon^{0})\,.
\end{align}
Here the $\hat{\mathcal{L}}_{(i)}$ are invariants of $\hat{g}^{(0)}_{ij}$ and its intrinsic curvature. The  logarithmic term is present only in even $d$, for the holographic Weyl anomaly \cite{Henningson:1998gx}. At any given $d$ only the terms that diverge as $\varepsilon\to 0$ are uniquely determined by the boundary metric. In dual terms, they are fixed by the definition of the theory in the ultraviolet, and are independent of the state of the CFT$_d$. We can rewrite them in terms of the (physical) metric induced at $\rho=\varepsilon$,
\begin{equation}\label{indmetric}
    	g_{ij} (x,\varepsilon) = \frac{\ell^2}{\varepsilon} \hat{g}_{ij} (x,\varepsilon) 
    	\,,
\end{equation}
which gives a divergent action of the form 
\begin{equation}\label{divact2}
    I_{\rm{div}} = \frac{1}{8\pi G_N} \int d^d x \sqrt{-g} \mathcal{L}\, , \quad \text{where}\quad  \mathcal{L}\equiv \mathcal{L}_{(0)} + \cdots + \mathcal{L}_{(\lceil d/2 \rceil - 1)} - \log (\varepsilon) \mathcal{L}_{(d/2)} \,,
\end{equation}
and where again the logarithmic term is present only in even $d$ (more about it below).\footnote{Notice that we have absorbed a factor of $\ell/2$ in $\mathcal{L}$, in order to match the conventions in \cite{Kraus:1999di}, which we will follow in the next subsection.}
The first three terms of $I_{\rm{div}}$ were presented in \eqref{divact}. 
Then, holographic renormalization is performed by adding a counterterm action $I_\text{ct}=-I_{\text{div}}$ to \eqref{bulkaction} in order to render the action finite when $\varepsilon\to 0$. The action that results is the quantum effective action of the CFT$_d$, and its variation with respect to the renormalized metric  $\hat{g}^{(0)}_{ij}$ generates the expectation value of the renormalized stress tensor of the CFT$_d$.
Adding to the action higher curvature terms that are finite when $\varepsilon\to 0$ corresponds to changing the renormalization scheme. 

 Our framework will, however, be different than that of holographic renormalization. Instead of regarding $\rho=\varepsilon$ as a regularization device to be eventually removed, we will keep it finite and non-zero, taking it to correspond to the location of a physical brane, and adding to the action \eqref{bulkaction} a purely tensional term for the brane
\begin{equation}\label{braneaction}
    I_{b} = - T \int_{\rho = \varepsilon} d^dx \sqrt{-g}\,.   
\end{equation}

Since the action is finite when $\varepsilon\neq 0$, no counterterms need to be added, and our theory will be completely well defined by \eqref{bulkaction} and \eqref{braneaction}, without any other boundary terms.\footnote{This is the case for de Sitter or Minkowski branes, but in Karch-Randall models infinite renormalization must still be performed. It will become clear that, for our purposes, we need not concern ourselves with this.} The expansion \eqref{divact2} can then be continued to arbitrarily high orders, producing additional densities $\mathcal{L}_{(n)}$ which depend on the metric on the brane $g_{ij}$ and its curvature. This expansion now includes terms that would not diverge when $\varepsilon\to 0$. Such terms are necessary in order to correctly reproduce the dynamics of the brane in the bulk, which is determined by the Israel junction conditions \cite{Israel:1966rt} derived from the brane action \eqref{braneaction}  \cite{deHaro:2000wj}. The infinite series of these terms constitute an effective gravitational action $I_{\rm{bgrav}}$ in $d$ dimensions, and the fact that the action \eqref{divact2} is large for small $\varepsilon$ reflects the strong localization of gravity on the brane.
In practice, one obtains all the gravitational terms $\mathcal{L}_{(n)}$ in $ I_{\rm{bgrav}}$ in a unique manner by deriving them for arbitrary $d$, without regard to whether they are finite or divergent in any specific dimension $d$, as we will do in the next subsection.

Now the entire action, when evaluated on a generic bulk solution, will be
\begin{equation}\label{actdec}
    I+I_b = I_{\rm{bgrav}}+I_{\rm{CFT}}\,.
\end{equation}
We can think of $I_{\rm{CFT}}$ as the finite-$\varepsilon$ counterpart of the bulk contribution that is not determined by the boundary metric, thus accounting for the state of the CFT$_d$, but some care must be exercised. The left-hand side of \eqref{actdec} is the action of a finite gravitational system with Einstein-Hilbert dynamics, plus a brane, in $d+1$ dimensions. The right-hand side recasts it in the form of a higher-curvature gravitational theory in $d$ dimensions, coupled to a cutoff CFT$_d$. This CFT$_d$ backreacts on the metric $g_{ij}$, so once the cutoff is introduced and the gravitational theory $I_{\rm{bgrav}}$ is defined, there is no more `renormalization scheme dependence' of the CFT$_d$.

Note that the effective action $I_{\rm{bgrav}}$ is unambiguously determined (up to total derivatives) by the exact theory that it is derived from. This is not typically the case with effective theories, which can be subject to field redefinitions that change their form. For instance, the metric in an effective gravitational theory may be redefined as $g_{ij}\to g_{ij} +\varepsilon \alpha R_{ij}+ \mathcal{O}(\varepsilon^2)$, with some arbitrary coefficient $\alpha$.\footnote{Field redefinitions that involve the conformal fields reduce to the previous ones by using the lower-order effective equations of motion.} However, in our case the metric $g_{ij} (x,\varepsilon)$ is exactly determined for finite $\varepsilon$ by its bulk definition \eqref{indmetric}, and moreover its dynamics is also exactly specified by the Israel junction conditions in the bulk. So the effective gravitational theory for $g_{ij} (x,\varepsilon)$ is free from such ambiguities. A minor subtlety remains in even $d$ for the anomaly term $\mathcal{L}_{(d/2)}$, which we will discuss in the next subsection.

Then, in \eqref{actdec}, the terms $I$, $I_b$ and $I_{\rm{bgrav}}$ are well defined, but the action of the CFT$_d$ is only specified through \eqref{actdec}\footnote{That is, unless we work in some specific version of AdS/CFT where the CFT is independently defined. We will not be assuming this.}. That is, the value of the CFT$_d$ action $I_{\rm{CFT}}$, and of any other magnitude derived from it (stress tensor, entropy, etc), is obtained as the difference between the bulk action $I+I_b$ and the  $d$-dimensional action $I_{\rm{bgrav}}$, when these are evaluated on any solution of the theory.

All of these considerations simply set the stage for our statement that, in this article, we will not be concerned with $I_{\rm{CFT}}$, but only with the gravitational action $I_{\rm{bgrav}}$. It is interpreted as the effective action of the gravity theory that is induced on the brane through the integration of the bulk degrees of freedom in the region $0\leq \rho \leq \varepsilon$. In dual terms, we integrate the ultraviolet degrees of freedom of the CFT$_{d}$ at energy scales above the cutoff.
Once we have obtained it this way, we study the effective gravitational theory on its own.

%%%%%%%%%%%%%%%%%%%%%%%%%%%%

\subsection{Algorithm for Counterterms}\label{subsec:algo}

The method of computing the effective action described previously is cumbersome, but there exist iterative algorithms that greatly simplify the calculations \cite{Kraus:1999di, Papadimitriou:2004ap, Elvang:2016tzz}.
Here we will follow \cite{Kraus:1999di}.

Let us define $\Pi_{ab}$ as the stress-energy tensor associated to the full effective action $I_{\rm bgrav}$, with Lagrangian $\mathcal{L} \equiv \mathcal{L}_{(0)} + \mathcal{L}_{(1)} + \cdots\,$,
\begin{equation}\label{DefPi}
    \Pi^{ab} \equiv \frac{2}{\sqrt{-g}}\frac{\delta}{\delta g_{ab}} \int d^d x \sqrt{-g} \, \mathcal{L}\,,
\end{equation}
and $\Pi$ as its trace, $\Pi \equiv g^{ab}\Pi_{ab}$.

The Gauss-Codazzi equations starting at the boundary are equivalent to the bulk Einstein equations in a Fefferman-Graham expansion.
The Gauss scalar constraint is
\begin{equation}\label{KLSeq}
    \frac{1}{d-1} \Pi^2 - \Pi_{ab} \Pi^{ab} = \frac{d(d-1)}{\ell^2} + R\,,
\end{equation}
where $R$ is the scalar curvature of the boundary metric $g_{ab}$. We will solve this equation order by order in the curvature, and then integrate \eqref{DefPi} to find the corresponding order-$n$ effective Lagrangian, $\mathcal{L}_{(n)}$. 

Two key observations were made in \cite{Kraus:1999di}.  First, one can start by taking 
\begin{equation}
    \Pi^{ab}_{(0)} = \frac{d-1}{\ell} g^{ab}\,,
\end{equation}
since at the leading order the terms that are proportional to the curvature can be neglected, implying that $\Pi^{ab}_{(0)}$ must be proportional to the metric. Second, by studying the behaviour of the counterterms under Weyl rescalings, one finds that the integration of \eqref{DefPi} must simply be
\begin{equation}\label{LPi}
    \mathcal{L}_{(n)} = \frac{1}{d-2n} \Pi_{(n)}\,,
\end{equation} up to total derivatives. This procedure then generates the corresponding order-$n$ term in the effective Lagrangian, and it can be iterated to compute the counterterms. We start from
\begin{equation}
    \Pi^{ab}_{(0)} = \frac{d-1}{\ell} g^{ab}, \qquad \Pi_{(0)} = \frac{d(d-1)}{\ell}, \qquad \mathcal{L}_{(0)} = \frac{d-1}{\ell}\,,
\end{equation}
and then we follow these steps iteratively:

\begin{enumerate}
    \item Knowing all $\Pi_{(i)}$ and $\Pi_{(i)}^{ab}$ of order less than $n$, solve for $\Pi_{(n)}$ using \eqref{KLSeq}.
    \item Compute $\mathcal{L}_{(n)}$ using \eqref{LPi}.
    \item Vary $\mathcal{L}_{(n)}$ to find $\Pi^{ab}_{(n)}$.
\end{enumerate}
In step 1, it is important to notice that at each order $n$, equation \eqref{KLSeq} involves terms of the form $\Pi_{(n)} \Pi_{(n-i)}$ and $\Pi_{ab}^{(i)} \Pi^{ab}_{(n-i)}$, with $i \leq n$. Since $\Pi_{(0)}^{ab}$ is proportional to $g^{ab}$, the term $\Pi_{ab}^{(0)} \Pi^{ab}_{(n)}$ is proportional to $\Pi_{(n)}$, and so indeed we find an equation for $\Pi_{(n)}$.
Moreover, for all orders $n \geq 2$, there are no other terms on the right-hand side of \eqref{KLSeq}, so we can directly solve for $\Pi_{(n)}$ to find
\begin{equation}\label{PiKLS}
    \Pi_{(n \geq 2)} = - \frac{\ell}{2} \sum_{i = 1}^{n-1} \left[ \frac{1}{d-1} \Pi_{(i)} \Pi_{(n-i)} - \Pi_{ab}^{(i)} \Pi^{ab}_{(n-i)} \right].
\end{equation}

Notice that when $d$ is even, the algorithm seems to break down for $n=d/2$ due to the divergence in \eqref{LPi}. The reason for this is the following. Even if, in our context, for $\varepsilon\neq 0$ the action $I+I_b$ is finite, when we expand it in powers of $\varepsilon$ there appears a logarithmic term. It reflects the fact that the integration of conformal degrees of freedom produces non-local terms, and in the effective theory it shows up as the trace anomaly \cite{Henningson:1998gx}. In the algorithmic approaches to the computation of counterterms,  it was shown in \cite{Papadimitriou:2004ap} that one must effectively replace $1/(d-2n) \to \log\varepsilon$. Therefore, in a braneworld construction where $\varepsilon$ is finite, the apparent divergence in $\mathcal{L}_{(d/2)}$ for even $d$ is an artifact.
A similar argument would also work for the divergences appearing in $\mathcal{L}_{(n)}$ for $n \geq d/2$.

For our purposes in this paper, we will not concern ourselves with these effects. The overall coefficients of each of the $\mathcal{L}_{(n)}$ terms will not play a role in our discussion, except in Sec.~\ref{BIse}, where we consider them in $d=3$ where there is no anomaly.

The iterative procedure explained above gives for the first terms, already presented in \cite{Kraus:1999di}, the result
\begin{align}
    \mathcal{L}_{(0)} & = \frac{d-1}{\ell} , \\
    \mathcal{L}_{(1)} & = \frac{\ell}{2(d-2)} R , \\ 
    \mathcal{L}_{(2)} & = \frac{\ell^3}{2(d-2)^2(d-4)} \left[ R_{ab}R^{ab} - \frac{d}{4(d-1)} R^2 \right] ,  \label{quada} \\
    \mathcal{L}_{(3)} & = - \frac{\ell^5}{(d-2)^3(d-4)(d-6)} \Bigg[ \frac{3d+2}{4(d-1)}RR_{ab}R^{ab} - \frac{d(d+2)}{16(d-1)^2}R^3 \nonumber \\
    & - 2R^{ab}R_{acbd}R^{cd} + \frac{d-2}{2(d-1)}R^{ab}\nabla_{a}\nabla_{b} R - R^{ab} \Box R_{ab} + \frac{1}{2(d-1)} R \Box R \Bigg] . \label{cubicca}
\end{align}
Since we are computing the brane effective action and not its counterterms, our results differ from those in \cite{Kraus:1999di} by an overall minus sign.

Using the \texttt{Mathematica} packages \texttt{xAct} \cite{xPerm:2008,Nutma:2013zea}, we have been able to extend these results to quartic and quintic order for general dimension $d$, and to sextic order for $d=3$.
For general dimension, the quartic term reads
\begin{align}
\mathcal{L}_{(4)} = - & \frac{\ell^7}{(d-2)^4(d-4)(d-6)(d-8)} \nonumber \\
\Bigg[ &  \frac{13 d^2 - 38 d - 80}{8 (d-1)(d-4)} R_{ab} R^{ab} R_{cd} R^{cd} + \frac{- 15 d^3 + 18 d^2 + 192 d + 64}{16 (d-4) (d-1)^2} R_{ab} R^{ab} R^2 \nonumber \\ 
& + \frac{d (5 d^3 + 10 d^2 - 112 d - 128)}{128 (d-4) (d-1)^3} R^4 + \frac{5 d^2 - 16 d - 24 }{(d-1)(d-4)} R^{ab} R^{cd} R R_{acbd} \nonumber \\
&- 12 R_{a}{}^{c} R^{ab} R^{de} R_{bdce} + 8 R^{ab} R^{cd} R_{ac}{}^{ef} R_{bdef} - 8 R^{ab} R^{cd} R_{a}{}^{e}{}_{c}{}^{f} R_{bedf}  \nonumber \\
&-\frac{2 (d-6)}{d-4}R^{ab} R^{cd} R_{a}{}^{e}{}_{b}{}^{f} R_{cedf} + \frac{d^2 + 4 d -36}{2 (d-4) (d-1)}  R_{bc} R^{bc} \nabla_{a}\nabla^{a}R  \nonumber \\ 
&+ \frac{- 7 d^2 + 22 d +32}{4 (d-4) (d-1)^2}R^2 \nabla_{a}\nabla^{a}R + \frac{4}{d-1} R^{bc} \nabla_{a}R_{bc} \nabla^{a}R - \frac{d+8 }{4 (d-1)^2} R \nabla_{a}R \nabla^{a}R \nonumber \\
& + \frac{3d-8}{d-1}R^{ab} \nabla_{a}R^{cd} \nabla_{b}R_{cd} + \frac{d(d-6) }{8 (d-4) (d-1)^2}\nabla_{a}\nabla^{a}R \nabla_{b}\nabla^{b}R \nonumber \\ 
& + \frac{1}{d-1} R \nabla_{b}\nabla^{b}\nabla_{a}\nabla^{a}R - \frac{(d-4)(d+2)}{4 (d-1)^2} R_{ab} \nabla^{a}R \nabla^{b}R + \frac{d-4}{d-1} R_{a}{}^{c} R_{bc} \nabla^{b}\nabla^{a}R \nonumber \\ 
&- \frac{5 d^3 - 38 d^2 + 64 d + 16}{4 (d-4) (d-1)^2}R_{ab} R \nabla^{b}\nabla^{a}R + \frac{3 d^2 - 20d + 28}{(d-1)(d-4)}R^{cd} R_{acbd} \nabla^{b}\nabla^{a}R \nonumber \\ 
&- \frac{(d-6)(d-2)^2}{8 (d-4) (d-1)^2} \nabla_{b}\nabla_{a}R \nabla^{b}\nabla^{a}R + \frac{d-4}{d-1} R^{bc}\nabla^{a}R \nabla_{c}R_{ab} - 8 R^{ab} \nabla_{e}R_{acbd} \nabla^{e}R^{cd}  \nonumber \\ 
&+ \frac{5 d^2 - 6 d - 64}{2 (d-1)(d-4)}R^{ab} R \nabla_{c}\nabla^{c}R_{ab} + \frac{(d-2)(d-6)}{2(d-1)(d-4)} \nabla^{b}\nabla^{a}R \nabla_{c}\nabla^{c}R_{ab} \nonumber \\ 
&+ \frac{(d-2)}{d-1} R_{ab} \nabla_{c}\nabla^{c}\nabla^{b}\nabla^{a}R + \frac{5 }{d-1}R \nabla_{c}R_{ab} \nabla^{c}R^{ab} + 12 R^{ab} R^{cd} \nabla_{d}\nabla_{b}R_{ac} \nonumber \\ 
&+ \frac{11d-6}{d-1} R^{ab} R^{cd} \nabla_{d}\nabla_{c}R_{ab} - \frac{d-6}{2 (d-4)}\nabla_{c}\nabla^{c}R^{ab} \nabla_{d}\nabla^{d}R_{ab} - 2 R^{ab} \nabla_{d}\nabla^{d}\nabla_{c}\nabla^{c}R_{ab} \nonumber \\ 
&- 4 R^{ab} \nabla_{b}R_{cd} \nabla^{d}R_{a}{}^{c} + 4 R^{ab} \nabla_{c}R_{bd} \nabla^{d}R_{a}{}^{c} + \frac{2(5d-22) }{d-4}R^{ab} R_{acbd} \nabla_{e}\nabla^{e}R^{cd} \Bigg].   \label{quaa}
\end{align}
The quintic and sextic terms are  too large to present here, and so we include them as a \texttt{Mathematica} ancillary file.

To finish, let us mention that the algorithm of \cite{Kraus:1999di} was improved in \cite{Papadimitriou:2004ap} into the dilatation operator method using a Hamiltonian formulation. This allowed to include matter fields, prove the equivalence of these algorithmic techniques to the holographic renormalization method of \cite{deHaro:2000vlm}, and rigorously recover the trace anomaly.
The method has been further explored \cite{Papadimitriou:2010as, Papadimitriou:2016yit}, and a practical implementation that circumvents the Hamiltonian framework has been presented in \cite{Elvang:2016tzz}.

\section{Holographic $c$-theorem and higher-curvature gravities}\label{sec:HoloCTheorem}

The theory of gravity $I_{\rm{bgrav}}$ that is induced on the brane may admit solutions that are asymptotically AdS, and indeed, this can always be achieved with a brane tension $T$ below a critical value. In this case, the theory may be thought of as putatively dual to a CFT$_{d-1}$ (at least at planar level). A necessary condition for this theory to be well defined  is that it satisfies a $c$-theorem. One of our goals is to show that, not only the CFT$_{d-1}$  dual to the theory $I_{\rm{bgrav}}$ satisfies this condition, but also that all the higher-curvature terms in this effective action separately do so.

In this section we review the holographic proof of the $c$-theorem, and the characterization of higher-curvature gravities which satisfy it. Most of the content here is a compilation of previous results, but we also make a few observations which do not seem to have appeared explicitly in the literature before.   

\subsection{RG flow geometry and $c$-function}

The holographic $c$-theorem  involves a domain-wall type ansatz 
\begin{equation}\label{dwa}
    ds^2= e^{2A(r)} \left[-dt^2+ d\textbf{x} {}^2 \right]+dr^2\, ,
\end{equation}
which, in the presence of a matter
%certain 
stress-energy tensor $T_{ab}$
satisfying the null energy condition (NEC), produces a profile for $A(r)$ which makes the solution interpolate between two asymptotically AdS$_{d}$ regions \cite{Freedman:1999gp,Girardello:1998pd}. From the dual CFT perspective, these correspond to UV and IR fixed points, where the metric function is asked to behave as 
\begin{equation}
    A(r \rightarrow +\infty)=\frac{r}{L_{{\rm AdS}_{\rm UV}}}\, , \qquad  A(r \rightarrow -\infty)=\frac{r}{L_{{\rm AdS}_{\rm IR}}}\, ,
\end{equation}
where $L_{{\rm AdS}_{\rm UV}}$, $L_{{\rm AdS}_{\rm IR}}$ characterize the AdS curvature radii at each end of the geometry. Since the central charge of a holographic CFT is in general proportional to a power of the AdS curvature radius measured in Planck units, these geometries appear to adequately represent holographic RG flows when going from $r\rightarrow +\infty$ to $r\rightarrow -\infty$.

The idea of the holographic $c$-theorem\footnote{Here we will use the term `$c$-theorem' to refer to monotonicity theorems in general dimensions, often called the `$c$-theorem', `F-theorem' and `a-theorem' in two-, three- and four-dimensional CFTs \cite{Zamolodchikov:1986gt,Casini:2012ei,Komargodski:2011vj,Casini:2017vbe}.} is then to construct a function $c(r)$---the RG monotone or `$c$-function'---which monotonously decreases along the flow. A weak version of the theorem would require that $c_{\rm UV} > c_{\rm IR}$, whereas a strong one (which we will aim for) demands monotonicity along the entire flow,
\begin{equation}
    c'(r) \geq 0 \quad \forall\, r\, .
\end{equation}

A natural way of constructing a candidate $c(r)$ is to find an expression for $c'(r)$ that is proportional to the combination $T_t^t-T_r^r$. Then, if the matter stress-tensor satisfies the NEC, this combination is negative semidefinite,
\begin{equation}
    T_t^t-T_r^r \overset{\rm \ssc NEC}{\leq} 0\, ,
\end{equation}
and hence any $c'(r)\propto -(T_t^t-T_r^r)$ with a non-negative proportionality constant does the job. In this article we will always assume that matter is minimally coupled to gravity, so that the NEC does not involve any curvature terms.

If we denote the equations of motion of a given higher-curvature theory with Lagrangian $\mathcal{L}$ by
\begin{equation}
    \mathcal{E}_{ab} \equiv \frac{1}{\sqrt{-g}}\frac{\delta}{\delta g_{ab}} \int d^d x \sqrt{-g} \ \mathcal{L},
\end{equation}
then the combination $\mathcal{E}_t^t-\mathcal{E}_r^r$ evaluated on \eqref{dwa} will in general be a complicated combination of terms involving $A(r)$ and its higher-order derivatives, making the identification of $c(r)$ cumbersome (or directly impossible).

An important simplification occurs for theories with equations of motion that become second-order in derivatives of $A(r)$ and are at most linear in $A''(r)$ when evaluated on \req{dwa}.  This condition can be most easily implemented, for general families of higher-curvature theories, at the level of the action \cite{Paulos:2010ke}. Indeed, let 
\begin{equation}
    I[A] = \int d^d x \sqrt{-g} \mathcal{L} \left[A\right]
\end{equation} be the on-shell action from the evaluation of the corresponding higher-curvature action on the metric \req{dwa}. It is easy to show that the Euler-Lagrange equation of $A(r)$ is proportional to the $tt$ component of the field equations evaluated on \eqref{dwa}, namely, 
\begin{equation}\label{ttEq}
    \frac{\delta I[A]}{\delta A}=-2(d-1) e^{(d-1)A(r)}\left.\mathcal{E}_t^t\right|_A\,.
\end{equation}
It follows that, whenever $I[A]$ is second-order in derivatives of $A(r)$ and linear in $A''(r)$, so is $\mathcal{E}_t^t$.

The additional independent equation, corresponding to $\mathcal{E}_r^r$, is related to $\mathcal{E}_t^t$ by the Bianchi identity
\begin{equation}\label{rrEq}
    \partial_r \left.\mathcal{E}_r^r\right|_A+(d-1)A'(r)\left.\mathcal{E}_r^r\right|_A=(d-1)A'(r)\left.\mathcal{E}_t^t\right|_A\, .
\end{equation}
This immediately implies that $\mathcal{E}_r^r$ does not contain terms involving derivatives of $A(r)$ higher than one (since it is the scalar constraint\footnote{The explicit form of the equation $\mathcal{E}_r^r$ can be obtained from the on-shell action of $d s^2= e^{2A(r)} \left[-d t^2+ d \bf{x}^2 \right]+N(r)^2d r^2$ by varying with respect to the lapse function $N(r)$ \cite{Arciniega:2018tnn}.}) and that the combination $\mathcal{E}_t^t-\mathcal{E}_r^r$ is second-order in derivatives and linear in $A''(r)$.    
Throughout the paper, when speaking about theories satisfying the holographic $c$-theorem, we will be referring to theories that satisfy these reduced-order properties.\footnote{These requirements are identical to the ones satisfied by higher-curvature gravities which produce generalized Friedman equations of second order for the scale factor when evaluated on a Friedmann-Lema\^itre-Robertson-Walker ansatz with flat spatial slices---see \eg \cite{Sinha:2010pm,Tekin:2016rdx,Arciniega:2018fxj,Arciniega:2018tnn,Cisterna:2018tgx}. } 

For theories of the above type, it is straightforward to construct a function $c(r)$ such that \cite{Freedman:1999gp,Myers:2010xs,Myers:2010tj} 
\begin{equation}
    c'(r)=-\frac{\pi^{\frac{d-3}{2}}}{8\Gamma\left[\tfrac{d-1}{2}\right] G_N }\, \frac{T_t^t-T_r^r}{A'(r)^{d-1}}\, ,
\end{equation}
where, as required, the right-hand side is positive semidefinite, including for even $d$ \cite{Myers:2010tj}. As observed in \cite{Sinha:2010ai,Myers:2010tj}, $c(r)$ can be obtained for these theories from the Wald-like \cite{Wald:1993nt} formula
\begin{equation}
    c(r)\equiv \frac{\pi^{\frac{d-1}{2}}}{2\Gamma[\tfrac{d-1}{2}] A'(r)^{d-2}}\frac{\partial \mathcal{L}}{\partial R^{tr}\,_{tr}}\, ,
\end{equation}
where the Lagrangian derivative components are evaluated on \req{dwa}. By construction, $c(r)$ coincides at the fixed points with the holographic central charges $c$.

\subsection{Constraints on theories}

When trying to construct theories that satisfy simple holographic $c$-theorems, Lovelock gravities \cite{Lovelock1,Lovelock2,Padmanabhan:2013xyr} are natural candidates, as they have second-order equations on general backgrounds. The $n$-th order Lovelock density is
\begin{equation}
    \mathcal{L}^{(n)}_{\rm Lovelock}\equiv \mathcal{X}_{2n}\equiv \frac{(2n)!}{2^n} \delta_{a_1}^{[b_1}\delta_{a_2}^{b_2}\cdots \delta_{a_{2n}}^{b_{2n}]} R^{a_1 a_2}_{b_1b_2} \cdots R^{a_{2n-1} a_{2n}}_{b_{2n-1} b_{2n}}\, .
\end{equation}
When $d$ is even, the density with $n=d/2$ is a topological invariant. All the higher order densities (with $n> (d-1)/2$ when $d$ is odd, and with $n> d/2$  when $d$ is even) vanish identically.  
Hence, Lovelock theories are too restricted to provide a non-trivial family of order-$n$ densities in arbitrary dimensions.

A different set can be obtained using the Schouten tensor
\begin{equation}
    S_{ab}=\frac{1}{d-2} \left[R_{ab}- \frac{1}{2(d-1)}g_{ab} R \right] 
\end{equation}
as a building block. The general relation
\begin{equation}
    R_{abcd}=C_{abcd} - 2(g_{a[c} S_{d]b}+g_{b[d}S_{c]a})
\end{equation}
and the fact that the Weyl tensor vanishes on the RG flow ansatz \req{dwa}, suggests considering the family \cite{Paulos:2012xe}
\begin{equation}\label{Pn}
    \mathcal{P}^{(n)}= \delta^{[b_1}_{a_1} \delta^{b_2}_{a_2}\cdots \delta^{b_n]}_{a_n} S^{a_1}_{b_1} \cdots S^{a_n}_{b_n} \, .
\end{equation}
This vanishes for $n>d$ because the totally antisymmetric product of Kronecker deltas is identically zero in that case,  but it has been shown that a simple limiting procedure\footnote{The idea involves computing $ \mathcal{P}^{(n)}$ for some $\bar d$ greater than the dimension of interest $d$, dividing by $(\bar d -d)$ and then taking the limit $\bar{d} \rightarrow d$ of the resulting expression.} can be applied to $ \mathcal{P}^{(n)}$, which gives non-trivial densities for additional orders and dimensions \cite{Alkac:2020zhg} (see also \cite{Gabadadze:2020tvt,Bergshoeff:2021tbz}).

One may also systematically consider all the densities of a given curvature order for fixed $d$, with arbitrary relative coefficients, and identify the combinations that satisfy the aforementioned conditions. At quadratic order, this selects the Gauss-Bonnet density 
\begin{equation}
    \mathcal{X}_{4}= R-4R_{ab}R^{ab}+R_{abcd}R^{abcd}
\end{equation}
and the Weyl-square term $C_{abcd}C^{abcd}$, which identically vanishes on \req{dwa}. The cubic case was studied in \cite{Myers:2010tj} for general $d$.  At that order there exist eight independent densities (there are fewer for low enough $d$), and the holographic $c$-theorem imposes two constraints on them, leaving six independent densities that satisfy all the requirements. 

Hence, in general, for fixed $d$ and $n$ there will be several independent densities  satisfying the holographic $c$-theorem.
However, it is natural to expect that the functional on-shell dependence on $A(r)$ for fixed $d$ and $n$ is unique---in particular, given $j$ order-$n$ densities satisfying the $c$-theorem, $\sum_j \alpha_j \mathcal{L}_j$, we would have
\begin{equation}
   \mathcal{E}_t^t\left|_A - \mathcal{E}_r^r\right|_A = \biggl(\sum_j c_j \alpha_j \biggr)\cdot F_n(A,A',A'')\, ,
\end{equation}
where the dependence on the gravitational couplings fully factorizes. This always allows us to change the basis of densities so that a single one of them contributes non-trivially to 
$\mathcal{E}_t^t - \mathcal{E}_r^r $, while all the others produce a vanishing contribution---\eg the Weyl-square density at quadratic order.

As for the explicit form of $\mathcal{E}_t^t$, $\mathcal{E}_r^r$ and   $F_n(A,A',A'')$ when evaluated on \req{dwa} for individual non-trivial densities, a quick inspection of various cases strongly suggests that these are always given by 
\begin{equation}
  \left.\mathcal{E}_t^t \right|_A \propto A'(r)^{2(n-1)}\left[(d-1)A'(r)^2+2nA''(r) \right]\, ,\quad \left.\mathcal{E}_r^r\right|_A \propto (d-1)A'(r)^{2n}\, ,
\end{equation}
up to an overall factor, and
\begin{equation}
    F_n(A,A',A'')= 2n A'(r)^{2(n-1)} A''(r)\, ,
\end{equation}
for general $n$ and $d$. The functional dependence of the $c$-function is then $c(r)\propto A'(r)^{2n-d}$.
This sort of `uniqueness' has been argued to hold for general curvature orders in $d=3$ in \cite{Paulos:2010ke}, and has been recently proven in \cite{Bueno:2022lhf}. In the same references, one can find a characterization of all the densities of any curvature order in $d=3$ that satisfy the $c$-theorem and which are constructed from general contractions of the metric and the Riemann tensor.

Several other properties have been observed to hold for gravities in three dimensions that satisfy a $c$-theorem.  At quadratic order, the resulting theory is the New Massive Gravity of \cite{Bergshoeff:2009hq}---more on this below.
At higher curvature orders, theories of this kind arise from an order-by-order expansion \cite{Gullu:2010pc,Gullu:2010st,Bueno:2022lhf}  of a Born-Infeld-type gravity \cite{Gullu:2010pc},  which in turn satisfies the holographic $c$-theorem  by itself \cite{Gullu:2010st,Alkac:2018whk}.  In addition, it has been found that certain theories that satisfy the holographic $c$-theorem---some of which involve explicit covariant derivatives---are equivalent to Chern-Simons gravities \cite{Afshar:2014ffa}. More recently, theories of this kind have been related to truncations of certain infinite-dimensional Lie algebras \cite{Bergshoeff:2021tbz}. It has also been shown that theories of this kind never propagate the scalar mode that is present in the linearized spectrum of generic higher-curvature theories \cite{Bueno:2022lhf}. This feature is likely valid for general $d$.

%%%%%%%%%%%%%%%%%%%%%%%%%%%%%%%%%%%%%%%%%%%%%%%%%%%%%%%%%%%%%%%%%%

\section{Holographic $c$-theorem for induced gravity}\label{sec:GenProof}

We will now prove one of our main results: all the densities in the action of holographically induced gravity, at arbitrary order $n$ and in general dimension $d$, belong in the class of theories whose dual CFTs satisfy a holographic $c$-theorem.

Before we proceed, let us emphasize that this is not the same as the monotonicity theorem---the $g$-theorem---for the theory that is dual to the brane in the doubly-holographic construction. The latter is dual to the entire system of the induced gravity on the brane plus the cutoff CFT$_d$ coupled to it. The holographic $g$-theorem proven for this system in \cite{Fujita:2011fp} amounts to showing that, as the brane moves deeper into the bulk, its curvature decreases---in CFT terms, flowing to the IR reduces the number of degrees of freedom that are dual to the brane.
This is not what we are doing. After deriving the induced gravitational action $I_{\rm{bgrav}}$, we take this theory on its own and disregard its coupling to the CFT$_d$. Then, our proof of a $c$-theorem for the putative dual CFT$_{d-1}$ is no longer related to the properties of the brane moving in the bulk.  

To prove the $c$-theorem we shall assume that our gravitational theory is coupled to a matter sector that satisfies the NEC and that this condition can be readily translated, via  the field equations, into a condition on the curvature terms as shown in the previous section. For this purpose we assume that matter is minimally coupled to gravity, so that no curvature terms enter the NEC. This assumption is consistent but technically unnatural, and it could be interesting to investigate if it can be relaxed.

That the entire theory $I_{\rm{bgrav}}$ might satisfy a holographic $c$-theorem might not be unexpected, given its origin in a `good' theory (Einstein-AdS in $d+1$ dimensions, plus a brane) but it is less obvious that the separate order-$n$ densities should also do it. 

We will give two proofs of this result, the first one applying an induction method to the algorithm described in Sec.~\ref{sec:Counterterms}, and the second one using the counterterms adapted for conformally flat boundaries obtained in \cite{Anastasiou:2020zwc}.

\paragraph{Inductive algorithm proof.}
An examination of the terms $\mathcal{L}_{(n)}$ obtained in Section \ref{sec:Counterterms}, evaluated on the RG-flow metric \eqref{dwa}, suggests that the following expression may be valid for general orders and dimensions,
\begin{equation}\label{LcThD}
    \mathcal{L}_{(n)}\big|_A = - C_n \frac{d-1}{d-2n} (A')^{2(n-1)} \left[ d (A')^2 + 2n A'' \right]\, , \quad \text{where} \quad C_n \equiv \ell^{2n-1} \frac{(2n - 3)!!}{(2n)!!}\, .
\end{equation}
Remarkably, this expression, if correct, directly implies that each and all of  the $\mathcal{L}_{(n)}$ satisfy a holographic $c$-theorem. We will now prove that \eqref{LcThD} is indeed correct.

We proceed by induction. We assume that \eqref{LcThD} is true for all orders $k < n$, and then we perform the algorithm of Section \ref{sec:Counterterms} to see that it is also valid for order $n$.

From \eqref{LPi}, the induction hypothesis implies that, for all $k < n$, we have
\begin{equation}\label{PicThD}
    \Pi_{(k)}\big|_A = - C_k (d-1) (A')^{2(k-1)} \left[ d (A')^2 + 2k A'' \right]\,.
\end{equation}
Then, following equations \eqref{ttEq} and \eqref{rrEq}, with $\mathcal{E}_{ab} = \Pi_{ab}/2$, we obtain  
\begin{align}
    \Pi_{(k)}^{tt} \ \big|_A & = - C_k e^{-2A} (A')^{2(n-1)} \left[ (d-1) (A')^2 + 2 n A'' \right] = - \Pi_{(k)}^{x_ix_i} \big|_A \\
    \Pi_{(k)}^{rr} \ \big|_A & = C_k (d-1) (A')^{2n}.
\end{align}
Now, using equations \eqref{LPi} and \eqref{PiKLS} we can compute  $\mathcal{L}_{(n)}\big|_A$. The result reads
\begin{align}
    \mathcal{L}_{(n)} \big|_A & = \frac{1}{d-2n} \Pi_{(n)} \big|_A \\
    & = - \frac{\ell}{2(d-2n)} \sum_{k = 1}^{n-1} \left[ \frac{1}{d-1} \Pi_{(k)} \Pi_{(n-k)} - \Pi_{ab}^{(k)} \Pi^{ab}_{(n-k)} \right]_A \\
    & = - \frac{d-1}{d-2n} (A')^{2(n-1)} \left[ d(A')^2 + 2n A'' \right] \frac{\ell}{2} \sum_{k = 1}^{n-1} C_k C_{n-k}\, .
\end{align}
Finally, using the identity
\begin{equation}
     \frac{\ell}{2} \sum_{k = 1}^{n-1} C_k C_{n-k} = \frac{\ell^{2n-1}}{2} \sum_{k = 1}^{n-1} \frac{(2k-3)!!(2(n-k)-3)!!}{(2k)!!(2(n-k))!!} = \ell^{2n-1}\frac{(2n-3)!!}{(2n)!!} = C_n \, ,
\end{equation}
it follows that $\mathcal{L}_{(n)} \big|_A$ indeed reduces to the form \req{LcThD}, which means that all the order-$n$ Lagrangians appearing in the effective action $I_{\rm{bgrav}}$ satisfy holographic $c$-theorems.

It would appear that the proof breaks down at $n = d/2$ for even $d$, but as discussed in Sec.~\ref{subsec:algo}, these divergences are easily avoided artifacts.

\paragraph{Proof with conformally flat counterterms.}
%The proof of the holographic c-theorem relies on the evaluation of each density $\mathcal{L}_{(n)}$ on the conformally flat metric \eqref{dwa} on the brane. 
Instead of computing the general brane effective action, and then evaluating it on the conformally flat metric \eqref{dwa}, we can directly compute the effective action for a conformally flat brane.
For this, we can use (minus) the counterterms for an AdS$_{d+1}$ bulk with a conformally flat boundary, recently obtained in \cite{Anastasiou:2020zwc}.\footnote{We are grateful to I.~Papadimitriou for bringing these results to our attention.} For $n \leq d/2$ these are
\begin{equation}\label{cflat}
    \mathcal{L}_{(n)} |_{\rm{c.flat}} = (-1)^{n} \ell^{2n-1} \frac{(2n-3)!!(d-n)!}{(d-2)!(d-2n)} \mathcal{P}^{(n)} ,
\end{equation}
where $\mathcal{P}^{(n)}$ is the product of Schouten tensors defined in \eqref{Pn}, along with the necessary dimensional regularization prescription for the $n = d/2$ term. 

Since we have seen in the previous section that the $\mathcal{P}^{(n)}$ satisfy the holographic $c$-theorem, \eqref{cflat} directly proves our result. Indeed, when evaluated on the metric \eqref{dwa}, the expression above coincides with \eqref{LcThD}, since
\begin{equation}
    \mathcal{P}^{(n)} |_{A} = \frac{(-1)^{n+1}} {(2n)!!} \frac{(d-1)!}{(d-n)!} (A')^{2(n-1)} \left[ d(A')^2 + 2n A'' \right] .
\end{equation}

For $n > d/2$, the limiting procedure of \cite{Alkac:2020zhg}, described in the previous section, gives non-trivial densities when applied to $\mathcal{P}^{(n)}$. When we evaluate these densities on \eqref{dwa}, they also match our results above.

%%%%%%%%%%%%%%%%%%%%

\section{Structure of counterterm densities in three dimensions}\label{sec:3DResults}
Now we take a closer look at the explicit structure of the densities $\mathcal{L}_{(n)}$ for $n\geq 2$.  We shall first study the case $d=3$, and then $d\geq 4$.

As argued in \eg \cite{Paulos:2010ke,Gurses:2011fv,Bueno:2022lhf}, in $d=3$ the most general higher-curvature density constructed from contractions of the metric and the Riemann tensor is a function of the three densities\footnote{In \cite{Paulos:2010ke}, the notation $\mathcal{R}_2$, $\mathcal{R}_3$ is used for the same contractions as in \req{densi} but with the Ricci tensor replaced by its traceless part.} 
\begin{equation}\label{densi}
  R\equiv g^{ab}R_{ab}\, , \quad  \mathcal{R}_2 \equiv R_{ab}R^{ab}\, , \quad \mathcal{R}_3 \equiv R_{a}^b R_{b}^c R_c^a\, .
\end{equation}
This follows from the fact that all Riemann curvatures are Ricci curvatures due to the vanishing of the Weyl tensor, along with the existence of Schouten identities which relate terms involving higher-order contractions of the Ricci tensor to the ones above.
In three dimensions, conformal flatness is equivalent to the vanishing of the Cotton tensor,
\begin{equation}
C_{abc}\equiv 2 \nabla_{[c} R_{a|b]}+ \frac{1}{2} \nabla_{[b|} R g_{a|c]}\,.
\end{equation}
Then, the metric \req{dwa} used for holographic RG flows has $C_{abc}=0$.
 
\subsection{Quadratic order}
As mentioned in the introduction, in $d=3$ the density $\mathcal{L}_{(n)}$ coincides, up to an overall factor, with the quadratic term in the New Massive Gravity \cite{Bergshoeff:2009hq}. This is given by
\begin{equation}
    \bar{\mathcal{L}}_2= R_{ab}R^{ab}-\frac{3}{8} R^2\,,
\end{equation}
where the overbar in $\bar{\mathcal{L}}$ simply indicates that we remove the overall factors containing $\ell$ from the expressions in \eqref{quada}--\eqref{quaa}.
NMG is known to satisfy a holographic $c$-theorem \cite{Sinha:2010ai}. An additional property of $\bar{\mathcal{L}}_2$ is that, when linearized around maximally symmetric backgrounds, it propagates no scalar mode. Moreover, the equations of motion of $\bar{\mathcal{L}}_2$ have second-order trace \cite{Oliva:2010zd}.

\subsection{Cubic order}
To cubic order, and up to an overall factor, \eqref{cubicca} gives
\begin{equation}
    \bar{\mathcal{L}}_3=\frac{11}{8} R \mathcal{R}_2 - \frac{15}{64 }R^3 - 2 R^{ac} R^{bd} R_{abcd}+\frac{1}{4} R^{ab} \nabla_a \nabla_b R- R_{ab} \Box R^{ab}+ \frac{1}{4} R \Box R \, .
\end{equation}
Integrating by parts and substituting the three-dimensional Riemann tensor in terms of Ricci tensors, this can be rewritten as
 \begin{equation}\label{cubicupto}
  \bar{\mathcal{L}}_3\overset{\nabla}{=}-\frac{29}{8} R \mathcal{R}_2 + 4 \mathcal{R}_3+\frac{49}{64} R^3+ \frac{3}{8} R \Box R - R_{ab} \Box R^{ab} \, ,
  \end{equation}
 where we have introduced the notation 
 \begin{equation}
     \overset{\nabla}{=}\quad:  \textrm{equal up to total derivatives}\,.
 \end{equation}
If we use that
\begin{align}
  C_{abc} C^{abc}
   & \overset{\nabla}{=} - 2R_{ab} \Box R^{ab} +\frac{3}{4} R \Box R + 6 \mathcal{R}_3 -5 R \mathcal{R}_2+R^3 \, ,
 \end{align} 
then \eqref{cubicupto} can be further rewritten as
\begin{equation}\label{cubicc}
     \bar{\mathcal{L}}_3 \overset{\nabla}{=} \mathcal{S}_3[R_{ab}] +T_3[C_{abc}\cdots,\nabla_a]\, , 
\end{equation}     
where
\begin{equation}
    \mathcal{S}_3[R_{ab}] \equiv \mathcal{R}_3 + \frac{17}{64} R^3 - \frac{9}{8} R \mathcal{R}_2\, ,
\end{equation}
and
\begin{equation}
T_3[C_{abc}\cdots,\nabla_a]\equiv \frac{1}{2} C_{abc} C^{abc}\, .
\end{equation}
On the one hand, $\mathcal{S}_3$ is the cubic generalization of NMG identified in \cite{Sinha:2010ai} as the most general density of that order ---not involving covariant derivatives of the Ricci tensor--- which satisfies a holographic $c$-theorem. 
On the other hand, $T_3$ involves explicit covariant derivatives of the Ricci tensor. However, since it is proportional to the Cotton tensor, which identically vanishes on \eqref{dwa}, it has no effect on the holographic RG flow. Then, $\bar{\mathcal{L}}_3$ satisfies a holographic $c$-theorem. 

As it turns out, this density has interesting additional properties. On the one hand, as observed in \cite{Afshar:2014ffa}, the criterion that cubic extensions of NMG do not propagate a scalar mode usually present in the spectrum of higher-curvature gravities, and that they admit a Chern-Simons formulation, restricts them to a general linear combination  of $\mathcal{S}_3$ and $T_3$. Hence, $\bar{\mathcal{L}}_3$ satisfies these two requirements---the first one is in fact implied when the holographic $c$-theorem is required, as shown in \cite{Bueno:2022lhf}.

On the other hand, $T_3$ had been previously singled out in \cite{Oliva:2010zd} using yet a different criterion: it is the cubic density with the lowest-order traced field equations in three dimensions. Indeed, it is the only cubic theory whose equations of motion have a trace which only contains terms involving up to three derivatives of the metric.\footnote{Note that $\mathcal{S}_3$ does not have equations of motion with a reduced-order trace, which means that the $c$-theorem property and the reduced-order trace one are not directly connected, even though there are cases in which they do coincide, such as NMG itself and the $T_3$ density.  }

\subsection{Quartic order}
At quartic order, evaluating \eqref{quaa} for $d=3$ gives
\begin{align}
\bar{\mathcal{L}}_4 =&
-\frac{83}{16}\mathcal{R}_2^2-17 R \mathcal{R}_3 +\frac{1155}{64} R^2 \mathcal{R}_2-\frac{3635}{1024} R^4 \
\nonumber \\ 
& -  \frac{31}{4} \mathcal{R}_2 \Box R + \frac{57}{16} R^2 \Box R - 10 R^{bc} \nabla_{a}R_{bc} \nabla^{a}R + \frac{53}{16} R \nabla_{a}R \nabla^{a}R \nonumber \\ 
& + \frac{1}{2} R^{ab} \nabla_{a}R^{cd} \nabla_{b}R_{cd} + \frac{9}{32} \Box R \Box R + \frac{1}{2} R \Box^2 R + \frac{5}{16} R_{ab} \nabla^{a}R \nabla^{b}R \nonumber \\ 
&-  \frac{11}{2} R_{a}{}^{c} R_{bc} \nabla^{b}\nabla^{a}R + \frac{61}{16} R_{ab} R \nabla^{b}\nabla^{a}R -  \frac{3}{32} \nabla_{b}\nabla_{a}R \nabla^{b}\nabla^{a}R \nonumber \\ 
&-  \frac{1}{2} R^{bc} \nabla^{a}R \nabla_{c}R_{ab}  -  \frac{47}{4} R^{ab} R \nabla_{c}\nabla^{c}R_{ab} + \frac{3}{4} \nabla^{b}\nabla^{a}R \Box R_{ab} + \frac{1}{2} R_{ab} \Box \nabla^{b}\nabla^{a}R \nonumber \
\\ 
& -  \frac{11}{2} R \nabla_{c}R_{ab} \nabla^{c}R^{ab} + 12 R^{ab} \
R^{cd} \nabla_{d}\nabla_{b}R_{ac} -  \frac{27}{2} R^{ab} R^{cd} \nabla_{d}\nabla_{c}R_{ab}  \nonumber \\ 
&-  \frac{3}{2} \Box R^{ab} \Box R_{ab} + 28 R_{a}{}^{c} R^{ab} \Box R_{bc} - 2 R^{ab} \Box^2 R_{ab} - 4 R^{ab} \nabla_{b}R_{cd} \nabla^{d}R_{a}{}^{c} \nonumber \\ 
&  + 4 R^{ab} \nabla_{c}R_{bd} \nabla^{d}R_{a}{}^{c} + 16 R^{ab} \nabla_{d}R_{bc} \nabla^{d}R_{a}{}^{c}\, .
\end{align}
Again, when we decompose it as
\begin{equation}
    \bar{\mathcal{L}}_4 \overset{\nabla}{=} \mathcal{S}_4[R_{ab}] +T_4[C_{abc}\cdots,\nabla_a]\,,
\end{equation}
where 
\begin{equation}
    \mathcal{S}_4[R_{ab}] \equiv \frac{5}{4} \mathcal{R}_3 R -\frac{15}{16} \mathcal{R}_2^2 - \frac{45}{64} R^2 \mathcal{R}_2 + \frac{205}{1024} R^4 \,,
    \end{equation}
and
\begin{align}
   & T_4[C_{abc}\cdots,\nabla_a]\equiv +R C_{abc}C^{abc}-\frac{11}{2}R_{b}^a C_{aef}C^{bef}+\frac{23}{4} R^ {ac}R^ {bd}\nabla_{a}C_{bcd} \notag \\ &-\frac{17}{2}R_{eb} R^{e}_c\nabla_a C^{bac}+\frac{5}{2}R R_{bc}\nabla_a C^{bac}-\frac{5}{4}C_{bcd}R^{ac}\nabla_a R^{bd}-\frac{11}{2} R_{c}^e C^{bac}\nabla_a R_{be}\,.
\end{align}
Similarly to the cubic case, we find that $\mathcal{S}_4$ is the quartic generalization of NMG---algebraic in the curvature---which non-trivially satisfies a holographic $c$-theorem \cite{Sinha:2010ai,Paulos:2010ke}. On the other hand, we see that $T_4$ is a linear combination of terms which always involve at least one Cotton tensor and therefore identically vanish when evaluated on the RG-flow metric \eqref{dwa}. Again, this makes evident that $\bar{\mathcal{L}}_4 $ satisfies a holographic $c$-theorem.

Motivated by the cubic case, we have tried to express $T_4$ as one of the theories identified in \cite{Afshar:2014ffa} by the criterion that they admit a Chern-Simons description, but we have not succeeded in doing so. It seems that such identification only works for the quadratic and cubic terms. Similarly, while  $T_3$ had the property of possessing a reduced order for the trace of its equations of motion, this is no longer the case for $T_4$, whose traced equations are of order six.

\subsection{Higher orders}

It seems, then, that of all the special properties that we identified for $\mathcal{S}_3$ and $T_3$ in $d=3$, only those that refer to the holographic $c$-theorem extend to higher orders. Of course we have already given a general proof that all the ${\mathcal{L}}_{(n)}$ satisfy this theorem, but we can aim at distinguishing a finer structure of how this happens.

We decompose the $\bar{\mathcal{L}}_n$ into terms $\mathcal{S}_n$ and $T_n$ such that the $\mathcal{S}_n$ contain all of the non-vanishing contribution to the $c$-function, and the $T_n$ vanish identically on the RG-flow metric \eqref{dwa}. For the lowest orders we have seen that this separation  can be performed in such a way that $\mathcal{S}_n$ is algebraic in the curvature, that is,
\begin{equation}\label{SnTn}
    \bar{\mathcal{L}}_n \overset{\nabla}{=} \mathcal{S}_n[R_{ab}] + T_n[\nabla_a, R_{ab}]\, .
    \end{equation}
In fact, in $d=3$ this decomposition can be performed in all $n$. This follows from the results in \cite{Paulos:2010ke,Bueno:2022lhf}, which show that, at every $n$, there always exists a density $\mathcal{C}_n[R_{ab}]$ which non-trivially satisfies the $c$-theorem.

For the cubic and quartic terms, we have found that the $T_n$ are proportional to the Cotton tensor. It is unclear whether this is the case also for the quintic term, since the expressions are exceedingly complicated.
On the other hand, the structure of $\mathcal{S}_n[R_{ab}]$ is uniquely constrained not only in $n=3,4$, as we have seen, but also in $n=5$. Up to that order, there is a single order-$n$ algebraic density $\mathcal{C}_n$ which non-trivially satisfies the holographic $c$-theorem \cite{Paulos:2010ke}, and so $\mathcal{S}_n[R_{ab}]$ must be proportional to it. The proportionality constant can be found by evaluating both $\bar{\mathcal{L}}_n$ and $\mathcal{C}_{n}$ on the RG-flow metric \eqref{dwa}. For the quintic case, we obtain $\mathcal{S}_5 = \frac{5}{64} \mathcal{C}_5$, where
\begin{equation}
    \mathcal{C}_{5} = \frac{61R^5}{960} - \frac{7R^3\mathcal{R}_2}{12} + \frac{2R^2\mathcal{R}_3}{15} + \frac{7R\mathcal{R}_2^2}{5} - \frac{16\mathcal{R}_2\mathcal{R}_3}{15}\, .
\end{equation}

However, degeneracies start to appear at order $6$. From that order on, there exist densities that are algebraic in the Ricci tensor and which trivially satisfy the holographic $c$-theorem \cite{Paulos:2010ke}. These have been characterized in a precise manner. As shown in \cite{Bueno:2022lhf}, there is a unique sextic density  of this type,\footnote{This is more easily written in terms of contractions of the traceless Ricci tensor $\tilde{\mathcal{R}}_2\equiv \tilde R_{a}^b \tilde R_b^a$, $\tilde{\mathcal{R}}_3\equiv \tilde R_{a}^b \tilde R_b^c \tilde R_c^a$, where $\tilde R_{ab}\equiv R_{ab}-\frac{1}{3}g_{ab}R$, namely, $\Omega_{(6)} = 6 \tilde{\mathcal{R}}_3 ^2-\tilde{\mathcal{R}}_2 ^3$. }
\begin{align}\label{Omeg6}
\Omega_{(6)} 
=\frac{1}{3}  \left[R^6- 9R^4\mathcal{R}_2+8R^3\mathcal{R}_3+21R^2\mathcal{R}_2^2-36R\mathcal{R}_2\mathcal{R}_3-3\mathcal{R}_2^3+18\mathcal{R}_3^2\right]\,,
\end{align}
with the important property that, at any order $n\geq 6$, all the densities algebraic in curvature that vanish on the RG flow geometry are proportional to $\Omega_{(6)}$. Then, by taking $L^{\rm general}_{n-6}$ to be the most general density that is algebraic in the curvature, we have that $L_{n-6}^{\rm general}\cdot \Omega_{(6)}$ is the most general density of that type at order $n$ that vanishes on  RG flows.

This implies that the characterization of the terms in \req{SnTn} is ambiguous for $n\geq 6$, since we can redefine
\begin{equation}
    \mathcal{S}_n' =\mathcal{S}_n + L_{n-6} \Omega_6\,,, \qquad T_n'= T_n-L_{n-6} \Omega_6
\end{equation}
where $L_{n-6}$ is an arbitrary order-$(n-6)$ density algebraic in the curvature. Still, it is possible that a particular separation exists such that $T_{n\geq6}$ does not involve any $\Omega_6$ and vanishes exclusively due to the presence of Cotton tensors in all its terms. If that is the case,  one can use this criterion to  give a unique definition for $\mathcal{S}_{n\geq 6}$.

As far as we know, there are two different proposals for special order-$n$ densities that non-trivially satisfy the holographic $c$-theorem. The first results from the expansion of the Born-Infeld-like extension of NMG presented in \cite{Gullu:2010pc}, and in the following subsection we find hints that this may indeed coincide with $\mathcal{S}_{n\geq 6}$ as defined by the above criterion. The second corresponds to a basis of densities selected by the fact that they satisfy a simple recursive formula which relates the order-$n$ representative to the order-$(n-1)$ and order-$(n-2)$ ones \cite{Bueno:2022lhf}.

\subsection{Born-Infeld gravities and counterterms}\label{BIse}

An interesting generalization of NMG with a Born-Infeld-type Lagrangian was proposed in \cite{Gullu:2010pc}. The Lagrangian is
\begin{equation}\label{BI-NMG}
    \mathcal{L}_{\text{BI-NMG}} =\alpha  \sqrt{\det \left( \delta_a^b + \beta  G_a^b \right) }\, ,
\end{equation}
where $G_{ab}$ is the Einstein tensor and $\alpha,\beta$ are constants.
This theory satisfies the holographic $c$-theorem \cite{Gullu:2010st}, and when expanded at low curvatures it also generates higher-derivative densities which non-trivially satisfy it at any truncated order \cite{Alkac:2018whk}. As we have seen, this property is shared by the effective gravitational action induced on the braneworld.

Following \cite{Bueno:2022lhf} we can expand $\mathcal{L}_{\text{BI-NMG}} $ order by order, to find higher-curvature densities $\mathcal{B}_{(n)}$ which, on the RG flow metric \eqref{dwa}, give
\begin{equation}
    \mathcal{B}_{(n)}[\alpha,\beta] \big|_A =\alpha \left( -\beta  \right)^n \frac{(2n-5)!!}{(2n)!!} (A')^{2(n-1)} \left[ 3(A')^2 + 2n A'' \right].
\end{equation}
Remarkably, if we take $\alpha=2/\ell$ and $\beta=-\ell^2$, then this result coincides, for all $n$,
with the RG flow of the order-$n$ braneworld density \eqref{LcThD} in $d=3$, namely
\begin{equation}
 \left. \mathcal{L}_{(n)}\right|_A  =  \mathcal{B}_{(n)}[2/\ell,-\ell^2] \big|_A \, .
\end{equation}
This result is highly non-trivial, since the coincidence occurs also for the relative factors between the different order-$n$ Lagrangians, and not only for the functional dependence in $A$ and its derivatives, which might have been expected. It is then natural to conjecture that the $d=3$ counterterm Lagrangian may be resummed as
\begin{equation}
    \mathcal{L} = \frac{2}{\ell} \sqrt{\det \left( \delta_a^b - \ell^2 G_a^b \right) } \ + \ \mathcal{T} [C_{abc}\cdots,\nabla_a],
\end{equation}
where again
\begin{equation}
    \mathcal{T} [C_{abc}\cdots,\nabla_a] \big|_A = 0.
\end{equation}
An even stronger conjecture would be that the whole tower of counterterms (including $ \mathcal{T} $) could be written as a Born-Infeld-like action. The idea
that Born-Infeld type actions may act as suitable AdS counterterms has been considered before in \cite{Mann:1999bt,Jatkar:2011ue,Sen:2012fc}.
%%%%%%%%%%%%%%%%%%%%%%%%%%%%%%%%%%%%%%%%%%%%%%%%%%%%%%%%%%%%%%%%%%

\section{Structure of counterterm densities in higher dimensions}
\label{sec:GenD}
Let us now move to $d\geq 4$. The expressions become considerably more involved than in three dimensions, but we can still infer a similar general structure based on the lowest orders.
For the following discussion, it will be useful to keep in mind that the Weyl tensor $C_{abcd}$ identically vanishes on the RG-flow geometry \req{dwa}.

\subsection{Quadratic order}
Up to an overall factor, the quadratic term reads 
\begin{equation}
    \bar{\mathcal{L}}_2= R_{ab}R^{ab}-\frac{d}{4(d-1)} R^2\,,
\end{equation}
which is the $d$-dimensional generalization of NMG. Since it can be rewritten as a linear combination of the Weyl tensor squared and the quadratic Lovelock density, namely,
\begin{equation}
    \bar{\mathcal{L}}_2 = \frac{d-2}{4(d-3)}\left[ C_{abcd}C^{abcd}-\mathcal{X}_4\right]\, ,
\end{equation}
it is easy to see why it also fulfills a holographic $c$-theorem. Similar to the $d=3$ case,  $\bar{\mathcal{L}}_2$ propagates no scalar mode when linearized around maximally symmetric backgrounds \cite{Hassan:2013pca,Bueno:2016ypa}. Moreover, $\bar{\mathcal{L}}_2$ also belongs to the set of quadratic theories which have the property of possessing equations of motion whose trace is second-order, since for $d\geq 4$, that set is given by an arbitrary linear combination of
$C_{abcd}C^{abcd}$ and 
the quadratic Lovelock density
$\mathcal{X}_4$
\cite{Nakasone:2009bn,Oliva:2010zd,Oliva:2010eb}.

\subsection{Cubic order}
The cubic density was written in \req{cubicca} above.
Observe first that integrating by parts this can be rewritten as
\begin{align}\label{L32}
    \bar{\mathcal{L}}_3\overset{\nabla}{=} &\notag +\frac{3d+2}{4(d-1)}R R_{ab}R^{ab}-\frac{d(d+2)}{16(d-1)^2}R^3-2 R^{ab}R_{acbd}R^{cd}\\ &-\frac{d}{4(d-1)} \nabla_a R \nabla^a R + \nabla^c R^{ab}\nabla_c R_{ab}\, .
\end{align}
Now, following inspiration from the three-dimensional case, we can try to rewrite $ \bar{\mathcal{L}}_3$ as a linear combination of densities with special properties. We find that, indeed, $ \bar{\mathcal{L}}_3$ can be written for general $d\geq 4$ as
\begin{equation}
\bar{\mathcal{L}}_3= \frac{d-2}{16(d-3)}\mathcal{N}_6+\Xi+\Delta \, ,
\end{equation}
where $\mathcal{N}_6$, $\Xi$ and $\Delta$ are distinguished for different reasons. On the one hand, $\mathcal{N}_6$, which is defined as
\begin{align}
    \mathcal{N}_6\equiv &-24 R^{abcd}R_{cdbe}R^{e}_a-\frac{3(d+2)}{d-1}R R^{abcd}R_{abcd}-\frac{24d}{d-2}R^{abcd}R_{ac}R_{bd}\\ \notag &-\frac{16d(d-1)}{(d-2)^2}R^{ab}R_{bc}R^c_a+\frac{12(d^3-2d^2+6d-8)}{(d-2)^2(d-1)}R R^{ab}R_{ab} \\ \notag &- \frac{d^4-3d^3+10d^2+4d-24}{(d-2)^2(d-1)^2}R^3\, ,
\end{align}
is the cubic Quasi-topological density \cite{Oliva:2010zd,Oliva:2010eb,Myers:2010ru}. This satisfies a number of interesting properties. Firstly, it can be written as 
\begin{equation}
   \mathcal{N}_6= \frac{d-2}{d-5} \left[4 W_1+8W_2-\mathcal{X}_6 \right]\, ,
\end{equation}
where $W_1\equiv C^{ab}{}_{cd}C^{cd}{}_{ef}C^{ef}{}_{ab}$, $W_2\equiv C_{abcd}C^{ebcf}C^a\,_{ef}\,^d$ and $\mathcal{X}_6$ is the cubic Lovelock density. This expression makes manifest that $ \mathcal{N}_6$ satisfies the holographic $c$-theorem \cite{Myers:2010tj}.  $ \mathcal{N}_6$ identically vanishes in $d= 4$ but it is non-trivial for $d\geq 5$. It is in fact the term involving $ \mathcal{N}_6$ (actually $\mathcal{X}_6$) the one which makes $\bar{\mathcal{L}}_3$ be non-trivial when evaluated on \req{dwa} for $d\geq 5$ ($d \geq 6$).  In addition, $ \mathcal{N}_6$ is one of the few cubic densities which possess second-order traced equations for general $d\geq 5$ \cite{Oliva:2010eb}.\footnote{For $d=5,6$ there are two independent densities which possess second-order traced equations whereas for $d\geq 7$ there exist three.} Finally, $ \mathcal{N}_6$ only propagates the usual massless graviton when linearized around maximally symmetric backgrounds and it admits particularly simple black hole solutions \cite{Oliva:2010eb,Myers:2010ru}.

On the other hand, $\Xi$ is the piece which contains the terms involving explicit covariant derivatives. It is explicitly given  by
\begin{equation}\label{sigom}
    \Xi\equiv \frac{(d-2)^2}{4(d-3)(d-6)}\left[ \Sigma + \frac{2(d-3)}{3(d-2)^2}\Theta\right]\, ,
\end{equation}
where $\Sigma$ and $\Theta$ were previously identified again in \cite{Oliva:2010zd} as the two only densities which possess field equations whose trace is third-order in derivatives for $d\geq 4$. They are given, respectively, by\footnote{Similarly to the case of $\mathcal{N}_6$ in $d=5$, the combination inside the brackets in \req{sigom} vanishes identically in $d=6$, and then one finds
\begin{align}\label{sigo9m}
    \Xi|_{d=6}\equiv &+\frac{2}{9} R^{abcd}R_{cdef}R^{ef}_{ab} - \frac{8}{9}R^{abcd}R_{ac}R_{bd}-\frac{4}{3}R^{abcd}R_{ac}R_{bd}+\frac{10}{9}R^{ab}R_{bc}R^c_a\\ \notag &+\frac{1}{450}R^3 -\frac{3}{10} \nabla_a R \nabla^a R+\nabla_a R_{bc}\nabla^a R^{bc}\, .
\end{align}}
\begin{align}
    \Sigma= & -\frac{3d-2}{2} R^{abcd}R_{cdef}R^{ef}_{ab} +\frac{8d}{3} R_{cd}^{ab}R_{bf}^{ce}R_{ae}^{df} +\frac{4d}{d-2}R^{abcd}R_{ac}R_{bd}\\ \notag &+ \frac{4(d-4)}{d-2}R^{ab}R_{bc}R^c_a-\frac{2d}{3(d-1)^2}R^3-\frac{d(d-3)}{(d-2)(d-1)}\nabla_a R \nabla^a R \\ \notag &+\frac{4(d-3)}{d-2}\nabla_a R_{bc}\nabla^a R^{bc} \, , 
    \end{align}
    and
    \begin{align}
    \Theta=& +2(d^2-4)R^{abcd}R_{cdef}R^{ef}_{ab}-4(d^2-4)R_{cd}^{ab}R_{bf}^{ce}R_{ae}^{df}-12(d-2)R^{abcd}R_{ac}R_{bd}\\ \notag &-16R^{ab}R_{bc}R^c_a+\frac{d^2-d+2}{(d-1)^2}R^3+\frac{6d}{d-1}\nabla_a R \nabla^a R-24\nabla_a R_{bc}\nabla^a R^{bc}   \, .
\end{align}
Both $\Sigma$ and $\Theta$ non-trivially fulfill the holographic $c$-theorem when evaluated on \req{dwa}. However, the combination appearing in the density $\Xi$ trivially satisfies the holographic $c$-theorem for general $d$, as it becomes a total derivative when evaluated on \req{dwa}.

Finally, $\Delta$ is a density which does not involve explicit covariant derivatives, which is trivial when evaluated on the holographic $c$-theorem ansatz for general $d$ and which does not satisfy any additional special property involving a reduced order for its traced equations. It is given by
\begin{align}
   \Delta &\equiv \frac{1}{d-3} \left[  \frac{(d-10)(d-2)}{24}R^{abcd}R_{cdef}R^{ef}_{ab}+\frac{36-d(10+7d)}{4(d-2)(d-1)}R R^{ab}R_{ab} \right. \\ \notag & +\frac{3(d-2)}{2} R^{abcd}R_{cdbe}R_a^e+\frac{3(d-2)(d+2)}{16(d-1)}R R^{abcd}R_{abcd}+\frac{d+8}{2}R^{abcd}R_{ac}R_{bd} \\ \notag & \left. +\frac{11d-16}{2}R^{ab}R_{bc}R^c_a+\frac{2(d-2)}{3}R_{cd}^{ab}R_{bf}^{ce}R_{ae}^{df} +\frac{-28+d(21d-16)}{24(d-2)(d-1)^2}R^3\right]\, .
\end{align}
As mentioned earlier,  the general set of cubic theories constructed from arbitrary contractions of the metric and the Riemann tensor satisfying the holographic $c$-theorem property was obtained in \cite{Myers:2010tj}. $ \Delta$ is one of the 5 independent densities which contribute trivially to the $c$-function.

In view of the three-dimensional case, it is natural to wonder whether all terms appearing in $\Xi$ and $\Delta$ may be rewritten in a simplified way in terms of the Weyl tensor---so that the fact that they vanish when evaluated on \req{dwa} becomes manifest.

An alternative decomposition of $\bar{\mathcal{L}}_{3}$, found in \cite{Anastasiou:2020zwc}, is
\begin{equation}
    \bar{\mathcal{L}}_{3} = S_{ab}\left(S_{cd} +  \frac{1}{d-3} \nabla_c \nabla_d \right)C^{acbd} + 3(d-4) \mathcal{P}^{(3)}.
\end{equation}
Since the Weyl tensor and $\mathcal{P}^{(3)}$ are explicit in this form, it makes manifest that $\bar{\mathcal{L}}_{3}$ satisfies the holographic $c$-theorem.

\subsection{Higher orders}
Going to higher orders complicates the expressions considerably. We presented the result for the general-$d$ quartic density in \req{quaa}. 
We have verified that, analogously to the $d=3$ case, it is also possible to write $\bar{\mathcal{L}}_4$ as a sum of a term which does not involve explicit covariant derivatives and which non-trivially satisfies the $c$-theorem, plus another one which does contain covariant derivatives and is trivial when evaluated on \req{dwa}. It is then natural to expect that the $n$-th order density in $d$ dimensions can always be written as
\begin{equation}
   \bar{ \mathcal{L}}_n= \mathcal{S}_n[R_{abcd}]+T_n[R_{abcd},\nabla_a]\,,
\end{equation}
where $\mathcal{S}_n[R_{abcd}]$ is linear in $A''(r)$ when evaluated on \req{dwa} and does not involve higher-derivative terms, and where $T_n[R_{abcd},\nabla_a]$ vanishes (or it is a total derivative) for the same ansatz.

%%%%%%%%%%%%%%%%%%%%%%%%%%%%%%%%%%%%%%%%%%%%%%%%%%%%%%%%%%%%%%%%%%

\section{Conclusions and Outlook}
Let us close with a few observations and possible future directions.

\paragraph{Computation of higher order counterterms.} We have implemented the algorithm of \cite{Kraus:1999di} in \texttt{Mathematica} to obtain the quartic and quintic counterterms for pure AdS$_{d+1}$ gravity. It would be interesting to see if the methods of \cite{Papadimitriou:2004ap, Elvang:2016tzz} allow easier computation of higher orders. Formulating the algorithm on a basis of Weyl and Schouten tensors may also reveal finer structures in the counterterms.

\paragraph{Higher-curvature gravities in the bulk.} We have seen that starting from Einstein gravity in the $(d+1)$-dimensional bulk, the effective $d$-dimensional  higher-curvature theories induced on the brane satisfy holographic $c$-theorems. What would happen if the bulk gravitational theory were itself a higher-curvature theory? It seems likely that the $c$-theorem we have proven is an imprint of the healthy dynamics of bulk Einstein gravity: {\it good parents raise good children}. In that case, we would expect it to fail for a general higher-curvature bulk theory. Natural exceptions to be expected are Lovelock gravities \cite{Lovelock1,Lovelock2}, which also have second order equations. In fact, it has been suggested in \cite{Brihaye:2008xu} that in that case the counterterm at a given order is a linear combination of the same Einstein gravity-induced counterterm plus a new piece proportional to the $d$-dimensional Lovelock density of the corresponding order. Hence, for instance, $\mathcal{L}_3$ would be a linear combination of \req{cubicca} plus the cubic Lovelock density $\mathcal{X}_6$, and so on. It would then follow that these modified brane actions also satisfy holographic $c$-theorems, since the Lovelock terms satisfy the required conditions---namely, second-order on-shell action and linearity in $A''(r)$ when evaluated on the \req{dwa} ansatz.
On a different front, it would be interesting to study possible implications or connections of the present results, both for Einstein gravity and higher-curvature bulk theories, to the `Kounterterm' holographic renormalization approach, which requires Weyl-flat boundaries and (in even $d$) vanishing Euler class  \cite{Olea:2006vd,Miskovic:2007mg,Anastasiou:2020zwc}. 

\paragraph{Counterterms as Born-Infeld gravities in higher-dimensions?} In Section \ref{BIse}, we showed that the order-$n$ counterterm Lagrangian $\mathcal{L}_{n}$ coincides, when evaluated on the holographic $c$-theorem metric ansatz \req{dwa}, with the general term resulting from the expansion of the Born-Infeld-type generalization of NMG \cite{Gullu:2010pc}. This suggests that the full three-dimensional counterterms Lagrangian might be rewritten in such a Born-Infeld form plus a possible term which would vanish when evaluated on the RG-ansatz metric \req{dwa}. A possible $d$-dimensional generalization of these observations is far from obvious at the moment, but a quick inspection of some low-dimensional cases suggests that the modified Born-Infeld-like Lagrangian 
\begin{equation}\label{BI-d}
    \mathcal{L}^{(d)}_{\text{BI}} =\alpha \left[\det \left( \delta_a^b + \beta  G_a^b \right) \right]^{\frac{1}{d-1}}
\end{equation}
also fulfills a simple holographic $c$-theorem. Moreover, when \req{BI-d} is evaluated on-shell (on \req{dwa}) and expanded order by order, we find densities $\mathcal{B}_{(n)}|_A$ with the same functional dependence on $A$ as in the on-shell counterterm Lagrangians \eqref{LcThD}. We have found, however, no straightforward way to define $\alpha$ and $\beta$ such that the relative (overall) coefficients match our findings in equation \eqref{LcThD}. It would be interesting to analyze this possibility in more detail and, more generally, to study the properties of the Lagrangian defined by \req{BI-d}.

\paragraph{Holographic $c$-theorem gravities and scalar modes.} We have seen that the counterterm Lagrangians of the lowest orders often satisfy additional properties besides the holographic $c$-theorem. One of them is the absence of the scalar mode that generically appears in the linearized spectrum around maximally symmetric backgrounds of higher-curvature theories---see \eg \cite{Bueno:2016ypa}. 
Many higher-curvature theories which satisfy the holographic $c$-theorem also seem to share this property. In fact, it has recently been proven in \cite{Bueno:2022lhf} that in $d=3$ all the higher-curvature theories that satisfy a holographic $c$-theorem propagate no scalar mode. 
It would be interesting to prove or disprove this for $d\geq 4$. Observe that the class of theories which do not propagate the scalar mode is larger than the class of theories that admit a holographic $c$-theorem, so the question is whether or not the latter class is fully contained within the former.  

In the case considered in this work, it seems natural that the higher-curvature gravities holographically induced on the brane should propagate no scalar mode when linearized around maximally symmetric backgrounds. This fact is true in $d=3$ to all orders, as we have just said, and in general $d$ at least for $n=2$. After all, these theories are induced from Einstein gravity in AdS$_{d+1}$. And from the bulk perspective and to linear order, it was shown already in \cite{Karch:2000ct} that one can choose an axial TT gauge for the (massless spin-2) $d+1$-dimensional graviton to induce an almost massless spin-2 $d$-dimensional graviton on the brane, plus an infinite tower of massive spin-2 modes.

On a similar note, it was recently shown in \cite{Hu:2022lxl} that the effective action of wedge holography (with two branes instead of one), which has the same structure as the brane effective action, could be described as a ghost-free multi-gravity. Again, the important point here was the fact that the bulk is Einstein gravity, so the brane effective action should not have ghosts. It could be interesting to investigate the absence of scalar modes in this approach.

%%%%%%%%%%%%%%%%%%%%%%%%%%%%%%%%%%%%%%%%%%%%%%%%%%%%%%%%%%%%%%%%%%

\section*{Acknowledgements}

We thank Pablo A.~Cano, Robie A.~Hennigar, Javier Moreno, Julio Oliva and Marija Tomašević for useful discussions, and Tomás Andrade for collaboration at early stages of the project. We also thank Ioannis Papadimitriou for clarifications on the algorithmical computation of the counterterms.
Most of the computations were performed using the \texttt{Mathematica} packages \texttt{xAct} \cite{xPerm:2008} and \texttt{xTras} \cite{Nutma:2013zea}, publicly available on \href{http://xact.es}{\texttt{http://xact.es}}.
The work of QL is  supported  by the Spanish Ministry of Universities through FPU grant No.  FPU19/04859.  RE and QL also acknowledge financial support from MICINN grant PID2019-105614GB-C22, AGAUR grant 2017-SGR 754, and State Research Agency of MICINN through the `Unit of Excellence María de Maeztu 2020-2023' award to the Institute of Cosmos Sciences (CEX2019-000918-M).

%%%%%%%%%%%%%%%%%%%%%%%%%%%%%%%%%%%%%%%%%%%%%%%%%%%%%%%%%%%%%%%%%%

\bibliographystyle{JHEP}
\bibliography{bib}

\end{document}